\documentclass[aps,preprint,unsortedaddress]{revtex4}%
\usepackage{amsfonts}
\usepackage{amsmath}
\usepackage{amssymb}
\usepackage{graphicx}%
\providecommand{\U}[1]{\protect\rule{.1in}{.1in}}

\begin{document}
\preprint{ }
\title[Microscopic mechanisms of the Fermi-liquid behavior]{Microscopic mechanisms for the Fermi-liquid behavior of Nb-doped strontium titanate}
\author{S. N. Klimin$^{a}$}
\author{J. Tempere$^{b}$}
\affiliation{Theorie van Kwantumsystemen en Complexe Systemen (TQC), Universiteit
Antwerpen, Universiteitsplein 1, B-2610 Antwerpen, Belgium}
\author{D. van der Marel}
\affiliation{D\'{e}partement de Physique de la Mati\`{e}re Condens\'{e}e, Universit\'{e} de
Gen\`{e}ve, CH-1211 Gen\`{e}ve 4, Switzerland}
\author{J. T. Devreese}
\affiliation{Theorie van Kwantumsystemen en Complexe Systemen (TQC), Universiteit
Antwerpen, Universiteitsplein 1, B-2610 Antwerpen, Belgium}

\begin{abstract}
The relaxation rate in Nb-doped strontium titanate involving different
scattering channels is investigated theoretically. It is demonstrated that the
total relaxation rate in SrTi$_{1-x}$Nb$_{x}$O$_{3}$ is provided mainly by two
mechanisms. The Baber electron-electron scattering with participation of both
Coulomb and phonon-mediated electron-electron interactions provides the
$T^{2}$-dependence of the relaxation rate. The scattering on the potential
landscape caused by impurities is responsible for the residual relaxation rate
at low temperatures. A good agreement with experiment is achieved accounting
for all phonon branches in strontium titanate, both the optical and acoustic
phonons. It is shown that the effective electron-electron interaction can be
attractive in strontium titanate, and provides superconductivity at low
temperatures and Fermi-liquid response in a wide range of temperatures. Thus
our microscopic model supports the notion that superconductivity and
Fermi-liquid properties of n-type SrTiO$_{3}$ have a common origin.

\end{abstract}
\date{\today}
\maketitle

\section{Introduction \label{sec:intro}}

In Ref. \cite{VDM2011}, a common origin of the Fermi-liquid properties and
superconductivity in Nb-doped strontium titanate has been postulated and
induced from the phenomenological treatment of the resistivity data and
superconducting critical temperatures $T_{c}$. In the present work, we deduce
the temperature dependence of the resistivity and the superconducting $T_{c}%
$'s from microscopic model calculations. The relaxation rate in SrTi$_{1-x}%
$Nb$_{x}$O$_{3}$ can be accounted for by various scattering mechanisms. The
Fermi-liquid properties of a charge carrier gas in a crystal are provided by
the effective electron-electron interaction. This interaction is an interplay
of the Coulomb repulsion and the phonon-mediated attraction provided by the
Fr\"{o}hlich electron-phonon interaction with the optical phonons. Strontium
titanate is a strongly polar crystal with static dielectric constant
$\varepsilon_{0}\approx240$ and high-frequency dielectric constant
$\varepsilon_{\infty}\approx5.44$ \cite{PRB-STO}. The ratio $\eta
=\varepsilon_{\infty}/\varepsilon_{0}$ is very small in strontium titanate.
Due to a small $\eta$, the Coulomb repulsion and the optical phonon-mediated
attraction significantly compensate each other. Thus the contribution of the
other electron-phonon interactions, e.g., the deformation interaction with the
acoustic phonons, can be important for the sign of the total effective
electron-electron interaction. The fact that n-doped SrTiO$_{3}$ is a
superconductor \cite{Koonce,Reyren} implies that in this material the
phonon-mediated attraction due to both optical and acoustic phonons can
overcome the Coulomb repulsion. This yields superconductivity in strontium
titanate at low temperatures about $T\sim1$ K. In this connection, we verify
the suggestion \cite{VDM2011} that the DC resistivity and the relaxation rate
in Nb doped SrTiO$_{3}$ can be due to the effective electron-electron
interaction. There are two channels for the Fermi-liquid response of metals
and strongly doped semiconductors with complex conductivity bands due to the
electron-electron interaction: the normal and Umklapp
\cite{Lawrence,McD1,McD2} scattering processes.

Several other scattering mechanisms besides the effective electron-electron
interaction are present in strontium titanate. For completeness, they must be
included in the treatment, as far as they may contribute to the total
resistivity. In the present work, the following additional scattering
mechanisms are considered.

(1) The electron-phonon interaction can contribute to the effective
electron-electron scattering as mentioned above. Besides this, the direct
scattering of the electrons by the LO phonons can contribute to the
resistivity of the polar crystals in the same way as to the optical absorption
\cite{PRB-STO,VDM-PRL,TD2001}.

(2) In the context of small polarons, the electron-phonon scattering results
in a temperature dependence of the resistivity which is close to the $T^{2}%
$-behavior \cite{Zhao}. Therefore the electron -- LO-phonon scattering may
influence the Fermi-liquid-like temperature dependence of the DC response of
the charge carriers.

As established in Refs. \cite{PRB-STO,VDM-PRL}, the polarons in strontium
titanate are large (Fr\"{o}hlich) polarons rather than the small polarons
treated in Ref. \cite{Zhao}. In this connection, the electron -- LO-phonon
scattering is considered here in terms of the large polarons.

(3) Finally, the scattering of the electrons on the potential landscape
induced by the impurities (for example, the Nb donors) can bring a
non-negligible contribution to the total relaxation rate and resistivity,
especially in the low-temperature regime.

\section{Theory and results \label{sec:theoryandresults}}

\subsection{Baber scattering}

Within a parabolic model for a simple conductivity band, the electron-electron
interaction due to the normal (i.e., within the first Brillouin zone)
scattering processes does not contribute to the carrier electric response
owing to momentum conservation, even for an anisotropic band \cite{Maebashi}.
However, for a non-parabolic and/or complex conductivity band with, e.g.,
light and heavy carriers, the carrier response can be non-zero as first found
by Baber \cite{Baber,Giamarchi}. Thus the resistivity can be provided by the
interaction between charge carriers with different band masses or between
electrons and holes.

The Baber scattering mechanism is the most promising candidate to provide the
main contribution to the total relaxation rate of Nb doped strontium titanate.
We calculate the resistivity using the Boltzmann equation. Further on, we
follow the notations of Ref. \cite{Ziman}. The non-equilibrium distribution
function $f_{n,\mathbf{k}}$ for the carriers in the $n$-th subband is
determined as%
\begin{equation}
f_{n,\mathbf{k}}=f_{n,\mathbf{k}}^{\left(  0\right)  }-\Phi_{n,\mathbf{k}%
}\frac{\partial f_{n,\mathbf{k}}^{\left(  0\right)  }}{\partial\varepsilon
_{n,\mathbf{k}}}, \label{a1}%
\end{equation}
where $f_{n,\mathbf{k}}^{\left(  0\right)  }\equiv f\left(  \varepsilon
_{n,\mathbf{k}}\right)  $ is the equilibrium electron distribution function in
the $n$-th subband of the conduction band,%
\begin{equation}
f\left(  \varepsilon_{n,\mathbf{k}}\right)  =\frac{1}{e^{\frac{\varepsilon
_{n,\mathbf{k}}-\mu}{k_{B}T}}+1},
\end{equation}
and $\varepsilon_{n,\mathbf{k}}$ is the electron energy.

The function $\Phi_{n,\mathbf{k}}$ is a measure of the deviation from
equilibrium in the electron distribution. The inner product of two real
functions $\Psi$ and $\Phi$ is defined by $\left\langle \Psi,\Phi\right\rangle
\equiv\sum_{n}\int d\mathbf{k}~\Psi_{n,\mathbf{k}}\Phi_{n,\mathbf{k}}$. The
Boltzmann equation can be written as
\begin{equation}
X=P\Phi, \label{BE}%
\end{equation}
where $P$ is the scattering operator which transforms the function $\Phi$ into
another function $\Psi=P\Phi$, and $X$ represents the left-hand side of the
Boltzmann equation,%
\begin{equation}
X=-\mathbf{v}_{n,\mathbf{k}}\cdot\frac{\partial f\left(  \varepsilon
_{n,\mathbf{k}}\right)  }{\partial T}\nabla T-\mathbf{v}_{n,\mathbf{k}}\cdot
e\frac{\partial f\left(  \varepsilon_{n,\mathbf{k}}\right)  }{\partial
\varepsilon_{n,\mathbf{k}}}\mathbf{E}. \label{X}%
\end{equation}
Here, $\mathbf{v}_{n,\mathbf{k}}$ is the velocity in the $n$-th subband,%
\begin{equation}
\mathbf{v}_{n,\mathbf{k}}=\frac{1}{\hbar}\frac{\partial\varepsilon
_{n,\mathbf{k}}}{\partial\mathbf{k}}. \label{vnk}%
\end{equation}

The collision integral of the Boltzmann equation can be represented as
$\left\langle \Phi,P\Phi\right\rangle $. The Boltzmann equation in the form
(\ref{BE}) implies that%
\begin{equation}
\left\langle \Phi,X\right\rangle =\left\langle \Phi,P\Phi\right\rangle .
\label{intf}%
\end{equation}
The variational principle established by Ziman \cite{Ziman} states that the
solution of the Boltzmann equation gives $\left\langle \Phi,P\Phi\right\rangle
$ its maximum value. Thus $\Phi_{n,\mathbf{k}}$ can be approximated by a trial
variational function.

In the notations of Ref. \cite{Ziman}, the DC resistivity is expressed by the
formula%
\begin{equation}
\rho=\frac{\left\langle \Phi,P\Phi\right\rangle }{\left[  \left\langle
\Phi,X\left(  E=1\right)  \right\rangle \right]  ^{2}}, \label{rho}%
\end{equation}
where $X\left(  E=1\right)  $ represents the left-hand side of the Boltzmann
equation in a unit electric field $E=1$ (and in the absence of the temperature
gradients). The normalization factor is determined by the expression%
\begin{equation}
\left\langle \Phi,X\left(  E=1\right)  \right\rangle =2\sum_{n}\int
e\mathbf{v}_{n,\mathbf{k}}\Phi_{n,\mathbf{k}}\frac{\partial f\left(
\varepsilon_{n,\mathbf{k}}\right)  }{\partial\varepsilon_{n,\mathbf{k}}}%
\frac{d\mathbf{k}}{\left(  2\pi\right)  ^{3}}. \label{norm}%
\end{equation}
The collision integral for the electron-electron scattering processes is%
\begin{align}
\left\langle \Phi,P\Phi\right\rangle  &  =\frac{1}{2k_{B}T}\sum_{n,n^{\prime}%
}\int\frac{d\mathbf{k}_{1}}{\left(  2\pi\right)  ^{3}}\frac{d\mathbf{k}_{2}%
}{\left(  2\pi\right)  ^{3}}\frac{d\mathbf{k}_{3}}{\left(  2\pi\right)  ^{3}%
}\frac{d\mathbf{k}_{4}}{\left(  2\pi\right)  ^{3}}\left(  \Phi_{n,1}%
+\Phi_{n^{\prime},2}-\Phi_{n,3}-\Phi_{n^{\prime},4}\right)  ^{2}\nonumber\\
&  \times\mathcal{P}\left(  n,1,n^{\prime},2\rightarrow n,3,n^{\prime
},4\right)  . \label{P1}%
\end{align}
Here, $\mathcal{P}\left(  n,1,n^{\prime},2\rightarrow n,3,n^{\prime},4\right)
$ is the scattering probability,%
\begin{align*}
\mathcal{P}\left(  n,1,n^{\prime},2\rightarrow n,3,n^{\prime},4\right)   &
=\left(  U_{\mathbf{k}_{1},\mathbf{k}_{3}}^{\left(  eff\right)  }\right)
^{2}f\left(  \varepsilon_{n,\mathbf{k}_{1}}\right)  f\left(  \varepsilon
_{n^{\prime},\mathbf{k}_{2}}\right)  \left[  1-f\left(  \varepsilon
_{n,\mathbf{k}_{3}}\right)  \right]  \left[  1-f\left(  \varepsilon
_{n^{\prime},\mathbf{k}_{4}}\right)  \right] \\
&  \times\frac{2\pi}{\hbar}\delta\left(  \varepsilon_{n,\mathbf{k}_{1}%
}+\varepsilon_{n^{\prime},\mathbf{k}_{2}}-\varepsilon_{n,\mathbf{k}_{3}%
}-\varepsilon_{n^{\prime},\mathbf{k}_{4}}\right) \\
&  \times\left(  2\pi\right)  ^{3}\delta\left(  \mathbf{k}_{1}+\mathbf{k}%
_{2}-\mathbf{k}_{3}-\mathbf{k}_{4}\right)  ,
\end{align*}
with $U_{\mathbf{k}_{1},\mathbf{k}_{3}}^{\left(  eff\right)  }$ the matrix
element of the effective electron-electron interaction, which includes both
the Coulomb repulsion and the phonon-mediated attraction.

The trial variational function $\Phi_{n,\mathbf{k}}$ is chosen in the form
\cite{Ziman}%
\[
\Phi_{n,\mathbf{k}}=\mathbf{v}_{n,\mathbf{k}}\cdot\mathbf{u}%
\]
where $\mathbf{u}$ is the unit vector parallel to the applied electric field.
Thus the collision integral is%
\begin{align}
\left\langle \Phi,P\Phi\right\rangle  &  =\frac{1}{2k_{B}T}\frac{1}{\left(
2\pi\right)  ^{9}}\frac{2\pi}{\hbar}\sum_{n,n^{\prime}}\int d\mathbf{k}%
_{1}d\mathbf{k}_{2}d\mathbf{k}_{3}d\mathbf{k}_{4}\left(  \left(
\mathbf{v}_{n,\mathbf{k}_{1}}+\mathbf{v}_{n^{\prime},\mathbf{k}_{2}%
}-\mathbf{v}_{n,\mathbf{k}_{3}}-\mathbf{v}_{n^{\prime},\mathbf{k}_{4}}\right)
\cdot\mathbf{u}\right)  ^{2}\nonumber\\
&  \times\left(  U_{\mathbf{k}_{1},\mathbf{k}_{3}}^{\left(  eff\right)
}\right)  ^{2}f\left(  \varepsilon_{n,\mathbf{k}_{1}}\right)  f\left(
\varepsilon_{n^{\prime},\mathbf{k}_{2}}\right)  \left[  1-f\left(
\varepsilon_{n,\mathbf{k}_{3}}\right)  \right]  \left[  1-f\left(
\varepsilon_{n^{\prime},\mathbf{k}_{4}}\right)  \right] \nonumber\\
&  \times\delta\left(  \varepsilon_{n,\mathbf{k}_{1}}+\varepsilon_{n^{\prime
},\mathbf{k}_{2}}-\varepsilon_{n,\mathbf{k}_{3}}-\varepsilon_{n^{\prime
},\mathbf{k}_{4}}\right)  \delta\left(  \mathbf{k}_{1}+\mathbf{k}%
_{2}-\mathbf{k}_{3}-\mathbf{k}_{4}\right)  . \label{P2}%
\end{align}

The electrons in the conduction band are described by the matrix Hamiltonian
from Ref. \cite{VDM2011}%
\begin{equation}
H=4\left(
\begin{array}
[c]{ccc}%
\varepsilon_{1}\left(  \mathbf{k}\right)  & 0 & 0\\
0 & \varepsilon_{2}\left(  \mathbf{k}\right)  & 0\\
0 & 0 & \varepsilon_{3}\left(  \mathbf{k}\right)
\end{array}
\right)  +\frac{1}{2}W, \label{H}%
\end{equation}
with the energies%
\begin{align}
\varepsilon_{1}  &  =t_{\delta}\sin^{2}\left(  \frac{a_{0}k_{x}}{2}\right)
+t_{\pi}\sin^{2}\left(  \frac{a_{0}k_{y}}{2}\right)  +t_{\pi}\sin^{2}\left(
\frac{a_{0}k_{z}}{2}\right)  ,\nonumber\\
\varepsilon_{2}  &  =t_{\pi}\sin^{2}\left(  \frac{a_{0}k_{x}}{2}\right)
+t_{\delta}\sin^{2}\left(  \frac{a_{0}k_{y}}{2}\right)  +t_{\pi}\sin
^{2}\left(  \frac{a_{0}k_{z}}{2}\right)  ,\nonumber\\
\varepsilon_{3}  &  =t_{\pi}\sin^{2}\left(  \frac{a_{0}k_{x}}{2}\right)
+t_{\pi}\sin^{2}\left(  \frac{a_{0}k_{y}}{2}\right)  +t_{\delta}\sin
^{2}\left(  \frac{a_{0}k_{z}}{2}\right)  ,
\end{align}
where $a_{0}$ is the lattice constant. The matrix $W$
\begin{equation}
W=\left(
\begin{array}
[c]{ccc}%
2D & \xi & \xi\\
\xi & 2D & \xi\\
\xi & \xi & -4D
\end{array}
\right)  \label{W}%
\end{equation}
describes the mixing of subbands within the conductivity band. We use the
values of the band parameters $t_{\delta},t_{\pi,}D,\xi$ from Ref.
\cite{VDM2011}: $t_{\delta}=35$ meV, $t_{\pi}=615$ meV, $\xi=18.8$ meV and
$D=2.2$ meV.

In the present treatment, we neglect the band mixing because it is relatively
small, and assume the anisotropic parabolic dispersion in each of three
subbands of the conductivity band (as in Refs. \cite{VDM2011,PRB-STO}) with
the tensor effective masses $\left\vert m_{jj}^{\left(  n\right)  }\right\vert
$ ($j=x,y,z,~n=1,2,3$). The masses $m_{jj}^{\left(  n\right)  }$ can take two
values: the \textquotedblleft light\textquotedblright\ mass $m_{\pi}%
=\frac{2\hbar^{2}}{a_{0}^{2}t_{\pi}}$ and the \textquotedblleft
heavy\textquotedblright\ mass $m_{\delta}=\frac{2\hbar^{2}}{a_{0}^{2}%
t_{\delta}}$. Using the above band parameters of strontium titanate and the
lattice constant $a_{0}=3.905%
\operatorname{\text{\AA}}%
$, we arrive at the band masses $m_{\pi}\approx1.6m_{e}$ and $m_{\delta
}\approx29m_{e}$, where $m_{e}$ is the electron mass in vacuum.

The effective electron-electron interaction in SrTiO$_{3}$ is considered in
the following way. In the standard diagram technique, the effective
interaction corresponding to an elementary electron-electron scattering
process for a single-branch electron-phonon system, is given by the expression
\cite{Mahan}%
\begin{equation}
U_{\mathbf{k},\mathbf{k}^{\prime}}^{\left(  eff\right)  }\equiv U_{e-e}%
^{\left(  eff\right)  }\left(  \mathbf{q},\omega\right)  =U_{C}\left(
\mathbf{q}\right)  +\frac{1}{\hbar}\left\vert V_{\mathbf{q}}\right\vert
^{2}\mathcal{D}^{\left(  0\right)  }\left(  \mathbf{q},\omega\right)
,\nonumber
\end{equation}
where $\mathcal{D}^{\left(  0\right)  }\left(  \mathbf{q},\omega\right)  $ is
the free-phonon Green's function,%
\begin{equation}
\mathcal{D}^{\left(  0\right)  }\left(  \mathbf{q},\omega\right)
=-\frac{2\omega_{\mathbf{q}}}{\omega_{\mathbf{q}}^{2}-\omega^{2}}, \label{D0}%
\end{equation}
$\mathbf{q}$ is the momentum transfer $\mathbf{q}=\mathbf{k}^{\prime
}-\mathbf{k}$, and $\omega$ is equal to the transition frequency
$\omega=\left(  \varepsilon_{n^{\prime},\mathbf{k}^{\prime}}-\varepsilon
_{n,\mathbf{k}}\right)  /\hbar$ (where $n,n^{\prime}$ are the indices of the
subbands of the conduction band). Thus the effective interaction is
\begin{equation}
U_{e-e}^{\left(  eff\right)  }\left(  \mathbf{q},\omega\right)  =U_{C}\left(
\mathbf{q}\right)  -\left\vert V_{\mathbf{q}}\right\vert ^{2}\frac
{2\hbar\omega_{\mathbf{q}}}{\hbar^{2}\omega_{\mathbf{q}}^{2}-\left(
\varepsilon_{n^{\prime},\mathbf{k}^{\prime}}-\varepsilon_{n,\mathbf{k}%
}\right)  ^{2}}. \label{ueff2}%
\end{equation}
The effective interaction (\ref{ueff2}) is written for a one-branch phonon
system and does not account for screening. For our calculations, both the
screening and the multi-branch phonon system must be accounted for. The
extension of the above formulae to the multi-branch phonon system is
straightforward:%
\begin{equation}
U_{e-e}^{\left(  eff\right)  }\left(  \mathbf{q},\omega\right)  =U_{C}\left(
\mathbf{q}\right)  -\frac{1}{\hbar}\sum_{\lambda}\frac{2\omega_{\mathbf{q}%
,\lambda}\left\vert V_{\mathbf{q},\lambda}\right\vert ^{2}}{\omega
_{\mathbf{q},\lambda}^{2}-\omega^{2}}-\frac{1}{\hbar}\frac{2\omega
_{\mathbf{q}}^{\left(  ac\right)  }\left\vert V_{\mathbf{q}}^{\left(
ac\right)  }\right\vert ^{2}}{\left(  \omega_{\mathbf{q}}^{\left(  ac\right)
}\right)  ^{2}-\omega^{2}}, \label{ueff3}%
\end{equation}
where $\lambda$ labels different optical-phonon branches. The contribution to
the effective interaction due to the acoustic phonons (with the interaction
amplitudes $V_{\mathbf{q}}^{\left(  ac\right)  }$) is written here as a
separate term. The screening for the Coulomb interaction and for the
phonon-mediated interaction can be introduced here as in Ref. \cite{Mahan},%
\begin{equation}
U_{e-e}^{\left(  eff\right)  }\left(  \mathbf{q},\omega\right)  =\frac{4\pi
e^{2}}{q^{2}\varepsilon_{\infty}\epsilon_{e}\left(  \mathbf{q},\omega\right)
}-\frac{1}{\hbar}\sum_{\lambda}\frac{\left\vert V_{\mathbf{q},\lambda
}\right\vert ^{2}}{\left[  \epsilon_{e}\left(  \mathbf{q},\omega\right)
\right]  ^{2}}\frac{2\omega_{\mathbf{q},\lambda}}{\omega_{\mathbf{q},\lambda
}^{2}-\omega^{2}}-\frac{1}{\hbar}\frac{2\omega_{\mathbf{q}}^{\left(
ac\right)  }\left\vert V_{\mathbf{q}}^{\left(  ac\right)  }\right\vert ^{2}%
}{\left(  \omega_{\mathbf{q}}^{\left(  ac\right)  }\right)  ^{2}-\omega^{2}},
\end{equation}
where $\epsilon_{e}\left(  \mathbf{q},\omega\right)  $ is the electron
screening factor. Note that screening is different for the Coulomb and
phonon-mediated interactions, and that the contribution due to the acoustic
phonons is not screened.

In the experimental situation of Ref. \cite{VDM2011}, the Fermi energy of the
electrons in the conduction band substantially exceeds their thermal energy
$k_{B}T$. Therefore only the electrons in a thermal layer near the Fermi
surface bring a dominating contribution to the relaxation rate. In other
words, the energies of the relevant electrons are close to the Fermi energy,
and the difference $\varepsilon_{n^{\prime},\mathbf{k}^{\prime}}%
-\varepsilon_{n,\mathbf{k}}\sim k_{B}T$ is relatively small with respect to
the Fermi energy.

The major contribution to the collision integral (\ref{P2}) lies in the range
of the momentum transfer $q\propto2k_{F}$, where $k_{F}$ is the electron wave
vector at the Fermi surface (in general, angle-dependent due to the band
anisotropy). In SrTiO$_{3}$, the corresponding acoustic-phonon energy
satisfies the condition $\hbar\omega_{\mathbf{q}}^{\left(  ac\right)  }\gtrsim
k_{B}T$ in the range $T<100%
\operatorname{K}%
$ for all samples treated in the experiment \cite{VDM2011} except the
lowest-doped sample, for which $\hbar\omega_{\mathbf{q}}^{\left(  ac\right)
}/k_{B}T\sim1$ at $T=100%
\operatorname{K}%
$. For the lower temperatures and/or for the higher doping contents the
aforesaid condition is fulfilled. The LO-phonon energies in SrTiO$_{3}$ are
much higher than the thermal energy in the considered temperature range. Thus,
owing to the fact that the electron-electron scattering effectively occurs
close to the Fermi surface, retardation effects in the phonon-mediated
interaction can be neglected. Therefore we can approximate the effective
interaction by the expression%
\begin{equation}
U_{e-e}^{\left(  eff\right)  }\left(  \mathbf{q}\right)  =\frac{4\pi e^{2}%
}{q^{2}\varepsilon_{\infty}\epsilon_{e}\left(  \mathbf{q}\right)  }%
-\sum_{\lambda}\frac{2\left\vert V_{\mathbf{q},\lambda}\right\vert ^{2}}%
{\hbar\omega_{\mathbf{q},\lambda}\left[  \epsilon_{e}\left(  \mathbf{q}%
\right)  \right]  ^{2}}-\frac{2\left\vert V_{\mathbf{q}}^{\left(  ac\right)
}\right\vert ^{2}}{\hbar\omega_{\mathbf{q}}^{\left(  ac\right)  }},
\label{Ueff5}%
\end{equation}
where $\epsilon_{e}\left(  \mathbf{q}\right)  $ is the static electron
screening factor. The different screening of the Coulomb and phonon-mediated
interactions, however, can strongly influence the resulting interaction. In
the present calculation, the Thomas-Fermi (TF) screening factor is used in the
effective interaction:%
\begin{equation}
\epsilon_{e}\left(  \mathbf{q}\right)  =1+\frac{\kappa_{s}^{2}}{q^{2}},
\label{TF}%
\end{equation}
where $\kappa_{s}$ is the inverse TF screening length,%
\begin{equation}
\kappa_{s}=\left(  \frac{6\pi e^{2}n_{0}}{\varepsilon_{\infty}E_{F}}\right)
^{1/2}, \label{xp}%
\end{equation}
$n_{0}$ is the carrier density, and $E_{F}$ is the Fermi energy. Note that
$\epsilon_{e}\left(  \mathbf{q}\right)  $ is the static screening factor for
the electron gas (without the lattice polarization). Therefore $\kappa_{s}$
contains the high-frequency dielectric constant $\varepsilon_{\infty}$ rather
than the static dielectric constant $\varepsilon_{0}$. The material parameters
for the Fr\"{o}hlich interaction with the optical phonons are taken the same
as in Ref. \cite{PRB-STO}. The values of the TO- and LO- phonon frequencies,
the actual electron densities and the plasma frequencies $\omega_{p}$
determined from the experiment \cite{VDM-PRL} are represented in Table 1.%

\begin{table}[h] \centering
\caption{Optical-phonon frequencies, electron densities and plasma frequencies of doped strontium titanate}%
\begin{tabular}
[c]{|l|l|l|l|l|l|l|l|l|}\hline
$x$ & $x=0.1\%$ & $x=0.1\%$ & $x=0.2\%$ & $x=0.2\%$ & $x=0.9\%$ & $x=0.9\%$ &
$x=2\%$ & $x=2\%$\\
$T$ & $T=7$ K & $T=300$ K & $T=7$ K & $T=300$ K & $T=7$ K & $T=300$ K & $T=7$
K & $T=300$ K\\\hline
$\hbar\omega_{TO,1}$ (meV) & 2.27 & 11.5 & 2.63 & 11.5 & 6.01 & 12.1 & 8.51 &
13.0\\\hline
$\hbar\omega_{LO,1}$ (meV) & 21.2 & 21.2 & 21.2 & 21.2 & 21.2 & 21.2 & 21.2 &
21.2\\\hline
$\hbar\omega_{TO,2}$ (meV) & 21.2 & 21.8 & 21.2 & 21.8 & 21.2 & 21.8 & 21.2 &
21.8\\\hline
$\hbar\omega_{LO,2}$ (meV) & 58.4 & 58.4 & 58.4 & 58.4 & 58.4 & 58.4 & 58.4 &
58.4\\\hline
$\hbar\omega_{TO,3}$ (meV) & 67.6 & 67.1 & 67.6 & 67.1 & 67.6 & 67.1 & 67.6 &
67.1\\\hline
$\hbar\omega_{LO,3}$ (meV) & 98.7 & 98.7 & 98.7 & 98.7 & 98.7 & 98.7 & 98.7 &
98.7\\\hline
$n_{0}$ (cm$^{-3}$) & $1.7\times10^{19}$ & $1.7\times10^{19}$ & $3.4\times
10^{19}$ & $3.4\times10^{19}$ & $1.5\times10^{20}$ & $1.5\times10^{20}$ &
$3.4\times10^{20}$ & $3.4\times10^{20}$\\\hline
$\hbar\omega_{p}$ (eV) & 0.1 & 0.1 & 0.14 & 0.14 & 0.29 & 0.29 & 0.44 &
0.44\\\hline
\end{tabular}
\label{table1}%
\end{table}%

For the acoustic-phonon contribution, we use the interaction amplitudes for
the deformation potential from Ref. \cite{PD1985}%
\begin{equation}
V_{\mathbf{q}}^{\left(  ac\right)  }=\left(  4\pi\alpha_{ac}\right)
^{1/2}\frac{\hbar^{2}}{m_{D}}q^{1/2} \label{Vqac}%
\end{equation}
with the dimensionless coupling constant%
\begin{equation}
\alpha_{ac}=\frac{E_{d}^{2}m_{D}^{2}}{8\pi n\hbar^{3}v}, \label{alphac}%
\end{equation}
where $n$ is the density, $E_{d}$ is the deformation potential, $v$ is the
sound velocity, and $m_{D}$ is the density-of-state band mass. It is i.a.
noting that the acoustic-phonon interaction amplitude (\ref{Vqac}) in fact
does not depend on the mass $m_{D}$. The sound velocity in strontium titanate
is taken $v\approx8.1\times10^{3}%
\operatorname{m}%
\operatorname{s}%
^{-1}$ \cite{Bell}.

The relaxation rate is determined using its relation to the resistivity,%
\begin{equation}
\frac{1}{\tau_{eff}}=\frac{\omega_{p}^{2}}{4\pi}\rho, \label{Tau}%
\end{equation}
where $\omega_{p}$ is the plasma frequency for the electron gas in the
conductivity band determined in the same way as in Ref. \cite{VDM2011}, and
listed in Table 1. In Fig. 1, we represent the numerical results for the
relaxation rate provided by Baber scattering as a function of temperature for
the actual electron densities, which are determined from the experimental
values of the doping level given in Ref. \cite{VDM2011}. We possess the
measured values for all material parameters of strontium titanate, except the
deformation potential $E_{d}$. In this connection, the value $E_{d}=23.3%
\operatorname{eV}%
$ has been chosen to fit the $T^{2}$-dependence of the relaxation rate to the
experimental data for the lowest doping level $x=0.1\%$. For the other
densities, we keep one and the same value of $E_{d}$. Typical values of the
deformation potential in other crystals found in the literature are about 10
to 30 eV, so that the chosen value is realistic.

The relaxation rates calculated with the same value of $E_{d}$ for all
densities reproduce the density dependence of the experimentally obtained
relaxation rate fairly well, except a residual background contribution which
does not vanish when the temperature tends to zero. Therefore the suggestion
\cite{VDM2011} that the $T^{2}$-dependence of the DC resistivity in
SrTi$_{1-x}$Nb$_{x}$O$_{3}$ can be provided by the Baber scattering mechanism
is supported by the present calculation.

When both optical and acoustic phonons are taken into account, the effective
electron-electron interaction at the Fermi surface can become attractive. With
optical phonons only (i.e., neglecting acoustic-phonon contribution) this
attraction is not possible. Moreover, the effective electron-electron
interaction without participation of the acoustic phonons is approximately
Coulomb-like. The relaxation rate for a Coulomb-like electron-electron
interaction monotonously decreases with an increasing density. The density
dependence of the relaxation rate observed experimentally can be explained by
a relative increase of the attraction provided by the acoustic phonons with
respect to the Coulomb repulsion and the interaction due to the optical
phonons. The latter two become more strongly screened when the density rises,
while the deformation potential is not screened. The increasing density
dependence of the relaxation rate therefore supports our hypothesis that the
phonon-mediated attraction (provided by both optical and acoustic phonons) in
strontium titanate can overcome the Coulomb repulsion even in the normal phase
-- at sufficiently high temperatures, when superconductivity does not exist.

\subsection{Other mechanisms}

We suggest that the relatively small residual relaxation rate in the Nb-doped
SrTiO$_{3}$ can be accounted for by the scattering of the electrons on the
potential landscape created by the impurities. The contribution to the DC
resistivity due to the scattering of the electrons by the potential of the
impurities is calculated using the approximation of the time-dependent
relaxation time within the Boltzmann equation approach. Assuming the
impurities chaotically distributed in space, the DC conductivity tensor is
determined by the expression (see Ref. \cite{Anselm})%
\begin{equation}
\sigma_{ij}=-\frac{2e^{2}}{\hbar^{3}}\sum_{n}\int\frac{d\mathbf{k}}{\left(
2\pi\right)  ^{3}}\tau_{n}\left(  \mathbf{k}\right)  \left(  \frac{\partial
f_{0}\left(  \varepsilon_{n,\mathbf{k}}\right)  }{\partial\varepsilon
_{n,\mathbf{k}}}\right)  \left(  \mathbf{v}_{n,\mathbf{k}}\right)  _{i}\left(
\mathbf{v}_{n,\mathbf{k}}\right)  _{j}%
\end{equation}
where $\left(  \mathbf{v}_{n,\mathbf{k}}\right)  _{j}$ are the components of
the electron velocity in the conduction band.

For the relaxation time $\tau_{n}\left(  \mathbf{k}\right)  $, we apply the
approximation of the isotropic threefold degenerate conduction band with the
effective density-of-states band mass $m_{D}\equiv\left(  m_{xx}^{\left(
n\right)  }m_{yy}^{\left(  n\right)  }m_{zz}^{\left(  n\right)  }\right)
^{1/3}$. In this approximation, $\tau_{n}\left(  \mathbf{k}\right)
=\tau\left(  \mathbf{k}\right)  $, and the conductivity tensor is reduced to
the scalar expression $\sigma\equiv\rho^{-1}$,%
\begin{equation}
\frac{1}{\rho}=-2\frac{e^{2}}{\hbar m_{D}^{2}}\int\frac{d\mathbf{k}}{\left(
2\pi\right)  ^{3}}k^{2}\tau\left(  \mathbf{k}\right)  \left.  \left(
\frac{\partial f_{0}\left(  \varepsilon\right)  }{\partial\varepsilon}\right)
\right\vert _{\varepsilon=\varepsilon\left(  \mathbf{k}\right)  },
\label{cond}%
\end{equation}
where $\rho$ is the DC resistivity. The momentum-dependent relaxation time
$\tau\left(  \mathbf{k}\right)  $ for the chaotically distributed impurities
with the density $n_{I}$ is determined using the differential scattering
cross-section $\sigma\left(  k,\theta\right)  $ as \cite{Anselm}
\begin{equation}
\frac{1}{\tau\left(  k\right)  }=2\pi n_{I}v_{\mathbf{k}}\int_{0}^{\pi}%
\sigma\left(  k,\theta\right)  \left(  1-\cos\theta\right)  d\theta
\label{tauk}%
\end{equation}
The scattering cross-section $\sigma\left(  k,\theta\right)  $ is calculated
here in the Born approximation \cite{Landau3}.%

\begin{equation}
\sigma\left(  k,\theta\right)  =\frac{m_{D}^{2}}{4\pi^{2}\hbar^{4}}\left\vert
\int U\left(  \mathbf{r}\right)  e^{-i\mathbf{q}\cdot\mathbf{r}}%
d\mathbf{r}\right\vert ^{2}, \label{Born}%
\end{equation}
where $U\left(  \mathbf{r}\right)  $ is the potential created by an impurity,
and $\mathbf{q}=\mathbf{k}^{\prime}-\mathbf{k}$ is the momentum transfer (with
$k^{\prime}=k$, because the scattering is elastic).

To the best of our knowledge, the true potentials created by the impurities in
doped strontium titanate are not yet reliably known. We suggest that the
majority contribution to the impurity scattering is due to the ionized niobium
donors. Thus we assume that the density $n_{I}\approx n_{0}$ (where $n_{0}$ is
the electron density), and that the potential created by an ion can be modeled
by a pseudopotential corresponding to the {\normalsize Yukawa (screened
Coulomb) }potential:%
\begin{equation}
U\left(  \mathbf{r}\right)  =-\frac{U_{0}}{r}\exp\left(  -\kappa_{I}r\right)
, \label{U}%
\end{equation}
where $U_{0}=e^{2}/\varepsilon_{0}$, $\kappa_{I}$ is the inverse screening
length for the impurity potential. The scattering cross-section $\sigma\left(
k,\theta\right)  $ with the potential (\ref{U}) is%
\begin{equation}
\sigma\left(  k,\theta\right)  =\left(  \frac{2m_{D}U_{0}}{\hbar^{2}}\right)
^{2}\frac{1}{\left(  2k^{2}\left(  1-\cos\theta\right)  +\kappa_{I}%
^{2}\right)  ^{2}}. \label{sigma}%
\end{equation}
Integrating in (\ref{tauk}) with (\ref{sigma}), the inverse momentum-dependent
relaxation time is obtained to be%
\begin{equation}
\frac{1}{\tau\left(  k\right)  }=\frac{2\pi^{2}n_{I}v_{\mathbf{k}}}{\kappa
_{I}}\left(  \frac{2m_{D}U_{0}}{\hbar^{2}}\right)  ^{2}\frac{1}{\left(
4k^{2}+\kappa_{I}^{2}\right)  ^{3/2}}. \label{tauk1}%
\end{equation}
The contribution to the total relaxation rate (\ref{cond}) provided by the
impurity scattering, $1/\tau_{eff}^{\left(  I\right)  }$, with (\ref{tauk1})
is calculated numerically.

Here, the parameter to be estimated is the inverse screening length
$\kappa_{I}$. First, we can calculate $\kappa_{I}$ using the Thomas-Fermi (TF)
approximation,%
\begin{equation}
\kappa_{I}^{\left(  TF\right)  }=\left(  \frac{6\pi n_{0}e^{2}}{\varepsilon
_{0}E_{F}}\right)  ^{1/2}, \label{TF1}%
\end{equation}
where $E_{F}=\frac{\hbar^{2}}{2m_{D}}\left(  \pi^{2}n_{0}\right)  ^{2/3}$ is
the Fermi energy for the threefold degenerate conduction band. We use here the
static (rather than high-frequency) dielectric constant because the ions do
not move, and hence the lattice polarization (provided by the polar LO
phonons) reduces the Coulomb potential by the ratio $\varepsilon_{\infty
}/\varepsilon_{0}$.

The results for the $1/\tau_{eff}^{\left(  I\right)  }$ with (\ref{TF}) are
shown in Fig. 2 (\emph{a}). The values of $1/\tau_{eff}^{\left(  I\right)  }$
in the zero-temperature limit are in fair agreement with the experiment
\cite{VDM2011} for the samples with two higher doping levels $x=0.9\%$ and
$x=2\%$. However, for the two weaker doped samples, the residual relaxation
rate calculated using (\ref{TF1}) is overestimated with respect to the
experiment. Therefore we can suggest that the spatial cutoff of the impurity
potential can be provided also by additional mechanisms which can be
irrelevant to the presence of the electron gas and which therefore do not
vanish at low electron densities. In this connection, the contribution to the
total relaxation rate due to the impurity scattering has been also calculated
using the trial screening radius $r_{I}\equiv\kappa_{I}^{-1}\approx0.76%
\operatorname{nm}%
$ which is one and the same for all samples and close to $\kappa_{I}^{\left(
TF\right)  }$ for the highest doping. The results for $1/\tau_{eff}^{\left(
I\right)  }$ in this case are shown in Fig. 2 (\emph{b}). We see that in this
case, the residual relaxation rate is in good agreement with the background
for the total relaxation rate measured in the experiment \cite{VDM2011}.
Moreover, the present result adequately reproduces the experimental density
dependence of the residual relaxation rate without varying $\kappa_{I}$.

In our treatment of the impurity screening length, we suppose this quantity to
be independent of doping in order to arrive at the experimentally observed
trend. The assumption that $\kappa_{I}$ does not depend on the doping means
that the dominant part of the impurity potential for the scattering processes
is essentially short-range. (Note that $r_{I}\approx0.76%
\operatorname{nm}%
$ is about twice the lattice parameter.) The Nb-ion is larger than the Ti-atom
and as a consequence the ions around the Nb-donors are radially displaced.
This has consequences for the local electronic structure and -- as far as we
know -- nobody has analyzed this theoretically for Nb-donors in SrTiO$_{3}$.
Empirically we know that even for a low Nb-concentration, no bound
donor-states are formed. This is totally different from the case of Si doped
with phosphor.

For the Yukawa potential (\ref{U}), a critical value $\kappa_{I}^{\left(
c\right)  }$ exists such that for $\kappa_{I}>\kappa_{I}^{\left(  c\right)  }%
$, there are no bound states in that potential. The Monte Carlo calculation
for a particle with an isotropic parabolic dispersion law \cite{Luo} gives
\begin{equation}
\tilde{\kappa}_{I}^{\left(  c\right)  }\approx1.1906\tilde{U},
\label{critical}%
\end{equation}
where $\tilde{\kappa}_{I}^{\left(  c\right)  }$ and $\tilde{U}$ are,
respectively, $\kappa_{I}^{\left(  c\right)  }$ and $U$ in the atomic units.
We can estimate $\tilde{\kappa}_{I}^{\left(  c\right)  }$ for an anisotropic
band using the parameters $\tilde{\kappa}_{I}^{\left(  c\right)  }$ and
$\tilde{U}$ expressed in the effective atomic units for a crystal:
$\tilde{\kappa}_{I}^{\left(  c\right)  }=\kappa_{I}^{\left(  c\right)  }%
a_{B}^{\ast}$, $\tilde{U}=U/\left(  a_{B}^{\ast}E_{h}^{\ast}\right)  $, with
the effective Bohr radius $a_{B}^{\ast}=\varepsilon_{0}\hbar^{2}/\left(
m_{D}e^{2}\right)  $ and the effective Hartree energy $E_{h}^{\ast}=m_{D}%
e^{4}\left(  \varepsilon_{0}\hbar\right)  ^{2}$. In Nb-doped strontium
titanate, the critical inverse screening radius corresponding to
(\ref{critical}), is $r_{I}^{\left(  c\right)  }\approx2.5%
\operatorname{nm}%
$. The aforesaid value $r_{I}\approx0.76%
\operatorname{nm}%
$ yields $\tilde{\kappa}_{I}\approx3.9\tilde{U}$, substantially larger than
$\tilde{\kappa}_{I}^{\left(  c\right)  }$. Thus, in accordance with the
aforesaid experimental evidence, there are no bound states in the Yukawa
potential with the parameters of the Nb-doped SrTiO$_{3}$ chosen in the
present calculation.

The other mechanisms which might contribute to the resistivity of SrTi$_{1-x}%
$Nb$_{x}$O$_{3}$ are the direct scattering of the electrons by the optical
phonons and the Umklapp electron-electron scattering. These contributions have
been separately calculated within the same kinetic equation approach as the
aforesaid contributions due to the Baber and impurity scattering. In Fig. 3,
the contribution to the relaxation rate due to the electron -- LO-phonon
scattering is plotted as a function of the temperature for the actual electron
densities in the samples studied experimentally in Ref. \cite{VDM2011}. Under
the experimental conditions of Ref. \cite{VDM2011}, the electron -- LO-phonon
scattering contribution to the relaxation rate appears to be relatively small
with respect to the relaxation rate provided by the Baber mechanism.
Furthermore, the relaxation rate due to the electron -- LO-phonon scattering
strongly decreases with decreasing temperature. Therefore the electron-phonon
scattering is not one of the dominating mechanisms of the DC conductivity in
Nb doped SrTiO$_{3}$.

Finally, the Umklapp electron-electron scattering was considered in Ref.
\cite{VDM2011} as one of possible sources for the DC resistivity in Nb-doped
SrTiO$_{3}$. For the electron resistivity in metals, the Baber mechanism plays
a minor role, and the Umklapp scattering can be sufficient to explain the
Fermi-liquid temperature dependence of the resistivity
\cite{Lawrence,McD1,McD2}. However, the Umklapp electron-electron scattering
remarkably contributes to the resistivity only when both the initial and final
momenta of the two electrons in the elementary scattering process lie in the
thermal layer near the Fermi surface. Therefore, e.g. in the simple case of a
spherically symmetric conductivity band, the Umklapp scattering does not
contribute to the resistivity when the maximal possible value of the momentum
transfer $q_{\max}=4\hbar k_{F}$ (where $k_{F}$ is the Fermi wave vector) is
smaller than the modulus of the reciprocal lattice vector $\mathbf{g}$.
Therefore the Umklapp processes can be non-negligible only at sufficiently
high densities. For an anisotropic conductivity band with a warped Fermi
surface (that is just the case in SrTiO$_{3}$) the restriction for the density
can be softened with respect to that for a spherically symmetric band.
However, it is not a priori known whether the Umklapp scattering is relevant
for the carrier densities in the experiment \cite{VDM2011}. Therefore we have
estimated the Umklapp contribution to the relaxation rate in strontium
titanate within the kinetic equation approach.

In Fig. 4, the relaxation rate provided by the Umklapp electron-electron
scattering (multiplied by $\hbar/\mathcal{G}^{2},$ where $\mathcal{G}<1$ is
the dimensionless interference factor for the Umklapp processes \cite{Ziman})
is plotted as a function of the electron density for the temperature $T=80$ K.
The relaxation rate provided by the Umklapp processes in strontium titanate is
not vanishingly small for densities higher than the threshold value
$n_{0}\approx8\times10^{20}%
\operatorname{cm}%
^{-3}$. However, the carrier densities achieved in the experiment
\cite{VDM2011} are substantially lower than this threshold value. In the
figure, the arrows show the actual electron densities in the experiment
\cite{VDM2011}. For the experimental densities $n_{0}\sim10^{19}$~cm$^{-3}$ to
$10^{20}$~cm$^{-3}$, the relaxation rate provided by the Umklapp
electron-electron scattering is vanishingly small with respect to the measured
values of $1/\tau$. The relaxation rate due to the Umklapp processes becomes
non-negligible in strontium titanate only for the electron densities
$n_{0}\gtrsim10^{22}%
\operatorname{cm}%
^{-3}$, much higher than the densities relevant for the experiment
\cite{VDM2011}. The conclusion follows that the contribution by Umklapp
processes to the Fermi-liquid behavior of relaxation rate and resistivity in
SrTi$_{1-x}$Nb$_{x}$O$_{3}$ is negligible. In summary, the electron --
LO-phonon and Umklapp scattering contributions to the DC resistivity of the Nb
doped strontium titanate bring only relatively small corrections in the
present treatment as far as it is related to the interpretation of the
experiment \cite{VDM2011}.

The relative contributions of different scattering mechanisms to the total
relaxation rate as a function of temperature are shown in Fig. 5 for the
sample with the doping content $x=0.002$. In the low-temperature range, the
residual scattering of the electrons on the potential landscape of the
impurities dominates. For higher temperatures, the Baber scattering plays a
key role providing the Fermi-liquid $T^{2}$-dependence of the total relaxation
rate. The electron -- LO-phonon scattering only slightly contributes to the
total relaxation rate and only at sufficiently high temperatures. Thus the
main contribution to the relaxation rate is brought by the Baber scattering at
moderate temperatures and by the scattering by the donors at low temperatures.

The total relaxation rate calculated taking into account the Baber,
electron-impurity, and electron -- LO-phonon scattering mechanisms is plotted
in Fig. 6. The $T^{2}$ temperature dependence of the relaxation rate is
completely due to the Baber mechanism. The scattering by donors is responsible
for the residual relaxation rate which constitutes the background resistivity
in SrTi$_{1-x}$Nb$_{x}$O$_{3}$. We see that the aforesaid mechanisms
convincingly explain both the temperature and density dependences for the
experimentally measured relaxation rate.

\section{Conclusions \label{sec:conclusions}}

In conclusion, the DC resistivity of Nb-doped strontium titanate is explained
in terms of two dominating mechanisms: Baber scattering with participation of
both Coulomb and phonon-mediated electron-electron interactions provides the
$T^{2}$-dependence of the resistivity and of the relaxation rate, while the
scattering on the potential landscape caused by impurities yields the residual
relaxation rate which does not vanish at $T=0$. The calculated relaxation
rates are in a good agreement with the experiment \cite{VDM2011}. Thus the
hypothesis on a common origin of two phenomena in SrTiO3 -- superconductivity
and the Fermi-liquid behavior of the resistivity -- is supported by the
microscopic calculations.

\begin{acknowledgments}
This work was supported by FWO-V projects G.0356.06, G.0370.09N, G.0180.09N,
G.0365.08, the WOG WO.035.04N (Belgium), the SNSF through Grant No.
200020-140761 and the National Center of Competence in Research (NCCR)
\textquotedblleft Materials with Novel Electronic Properties-MaNEP".
\end{acknowledgments}

\newpage%

\begin{figure}
[ph]
\begin{center}
\includegraphics[
height=4.6337in,
width=6.0277in
]%
{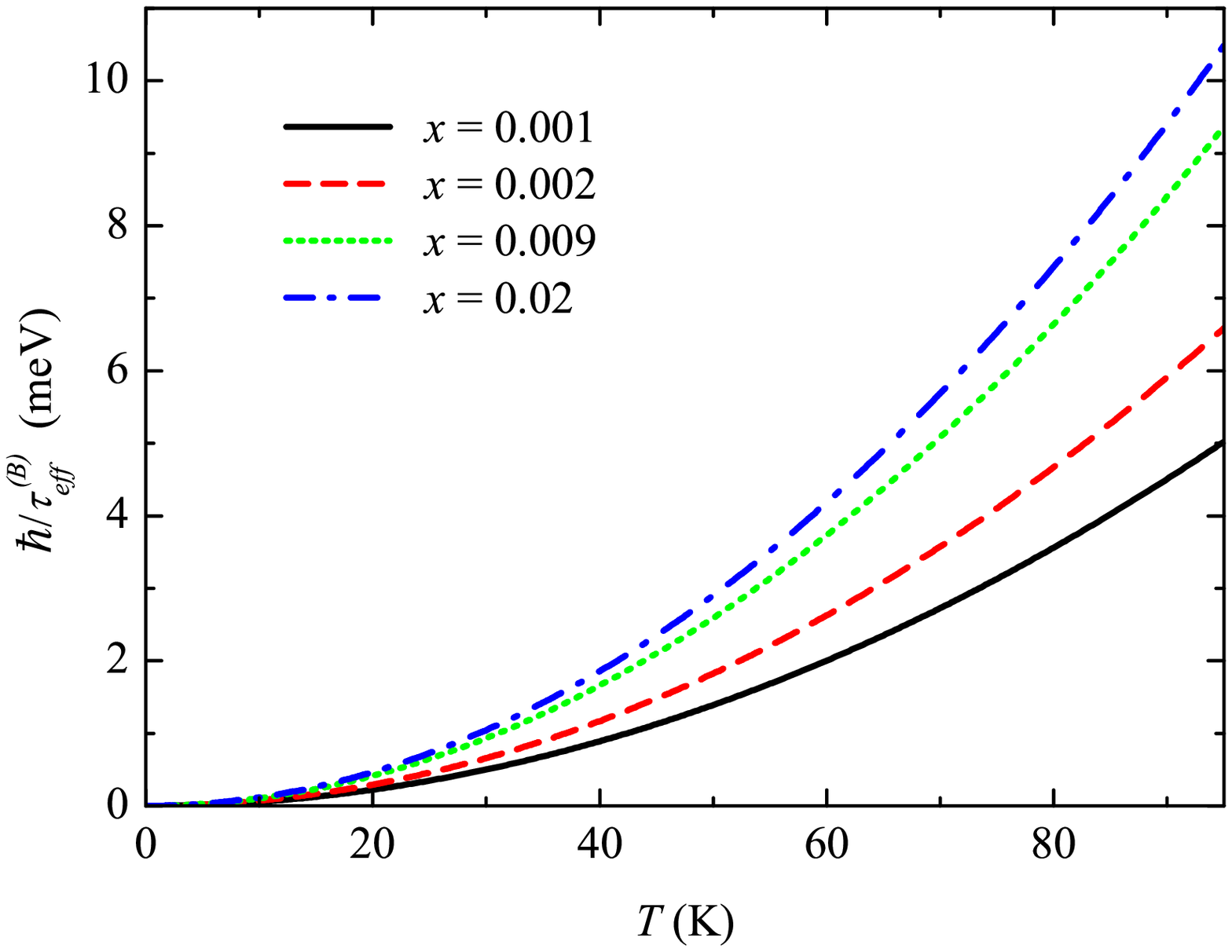}%
\caption{Relaxation rate provided by the Baber scattering in Nb-doped
SrTiO$_{3}$ as a function of the temperature for the actual electron densities
corresponding to the experimental values of the doping \cite{VDM2011}.}%
\end{center}
\end{figure}

\newpage%

\begin{figure}
[ph]
\begin{center}
\includegraphics[
height=7.632in,
width=5.7588in
]%
{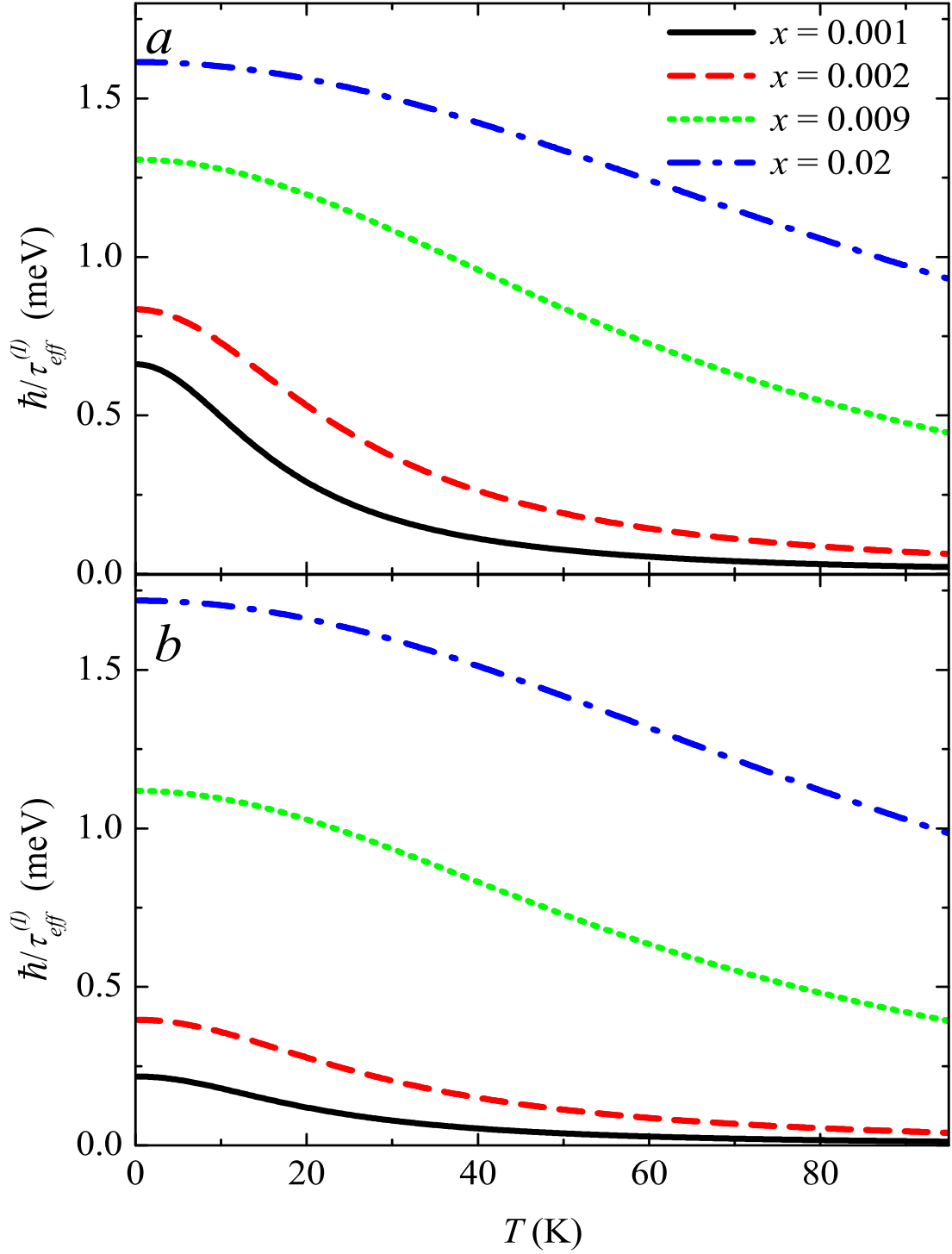}%
\caption{Residual relaxation rate provided by the scattering of the electrons
on the impurities in Nb-doped SrTiO$_{3}$ as a function of the temperature
(\emph{a}) using the Thomas-Fermi screening length, (\emph{b}) using the model
screening length, the same for all samples.}%
\end{center}
\end{figure}

\newpage%

\begin{figure}
[ph]
\begin{center}
\includegraphics[
height=4.7219in,
width=6.1661in
]%
{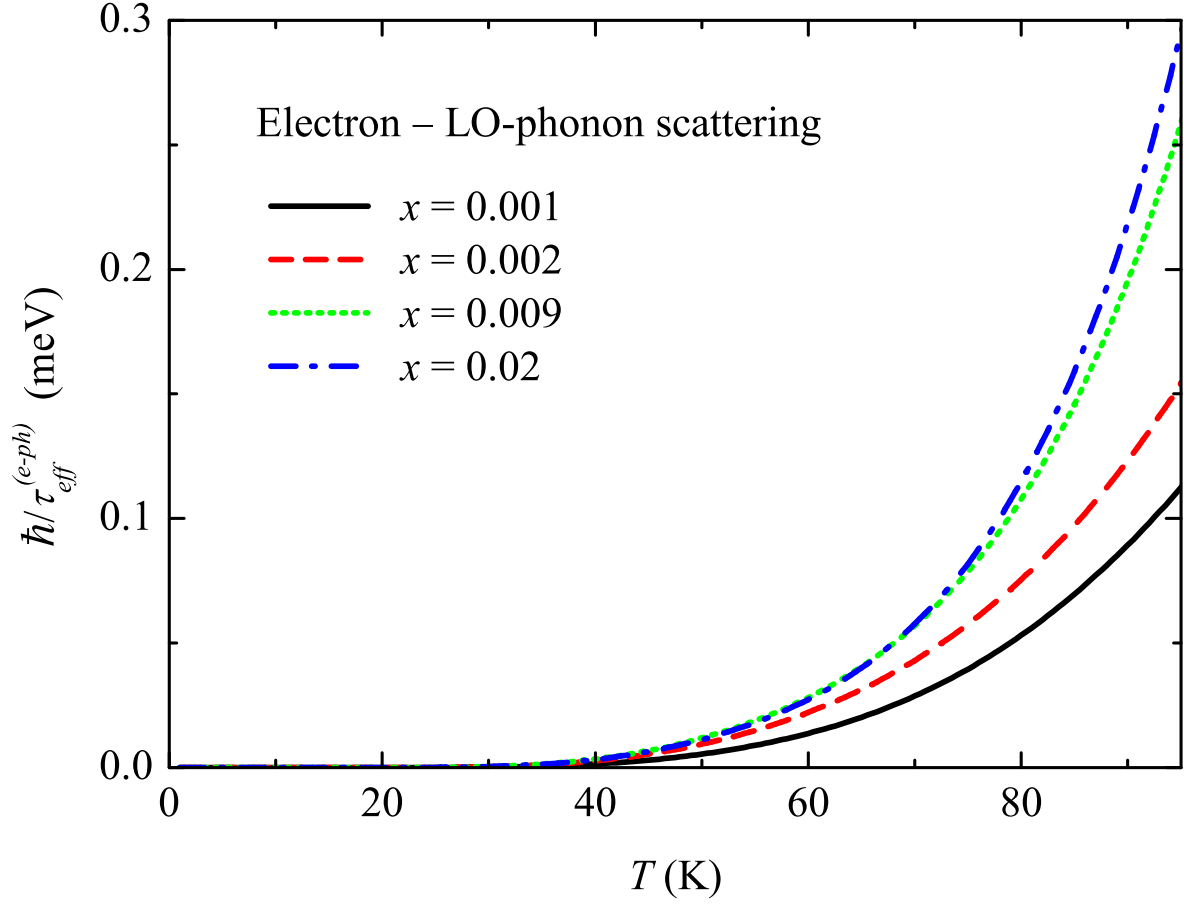}%
\caption{Contribution to the relaxation rate in Nb-doped SrTiO$_{3}$ due to
the direct electron -- LO-phonon scattering as a function of the temperature
for the electron densities from Ref. \cite{VDM2011}.}%
\end{center}
\end{figure}

\newpage%

\begin{figure}
[ph]
\begin{center}
\includegraphics[
height=4.6972in,
width=6.1012in
]%
{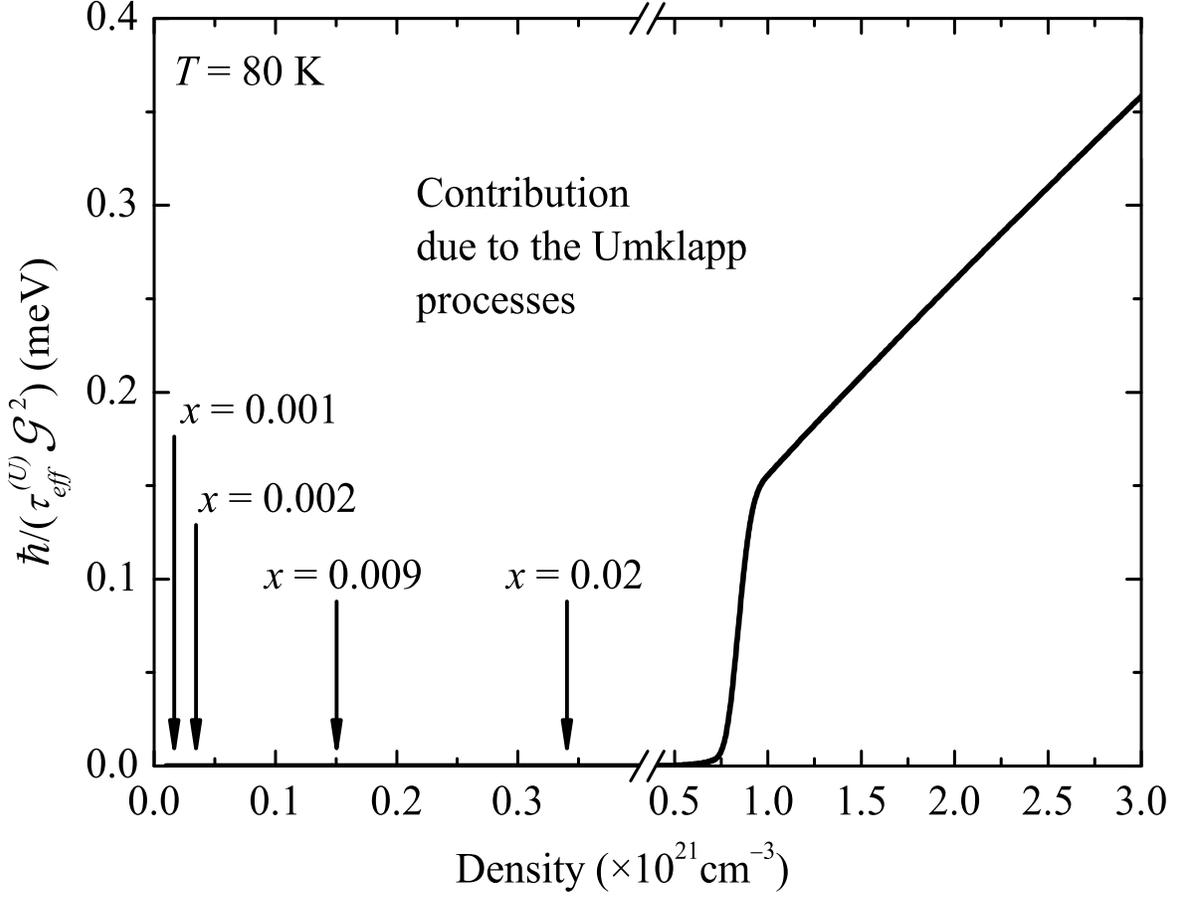}%
\caption{Relaxation rate due to the Umklapp electron-electron scattering in
SrTi$_{1-x}$Nb$_{x}$O$_{3}$ multiplied by $\hbar/\mathcal{G}^{2}$ (where
$\mathcal{G}$ is the interference factor for the Umklapp processes
\cite{Ziman}) as a function of the electron density for $T=80$ K. The arrows
indicate the electron density for the samples of the experiment \cite{VDM2011}%
. There are different scales at the $x$ axis for the densities $n_{0}%
<4\times10^{20}\operatorname{cm}^{-3}$ and $n_{0}>4\times10^{20}%
\operatorname{cm}^{-3}$.}%
\end{center}
\end{figure}

\newpage%

\begin{figure}
[ph]
\begin{center}
\includegraphics[
height=5.0341in,
width=6.3451in
]%
{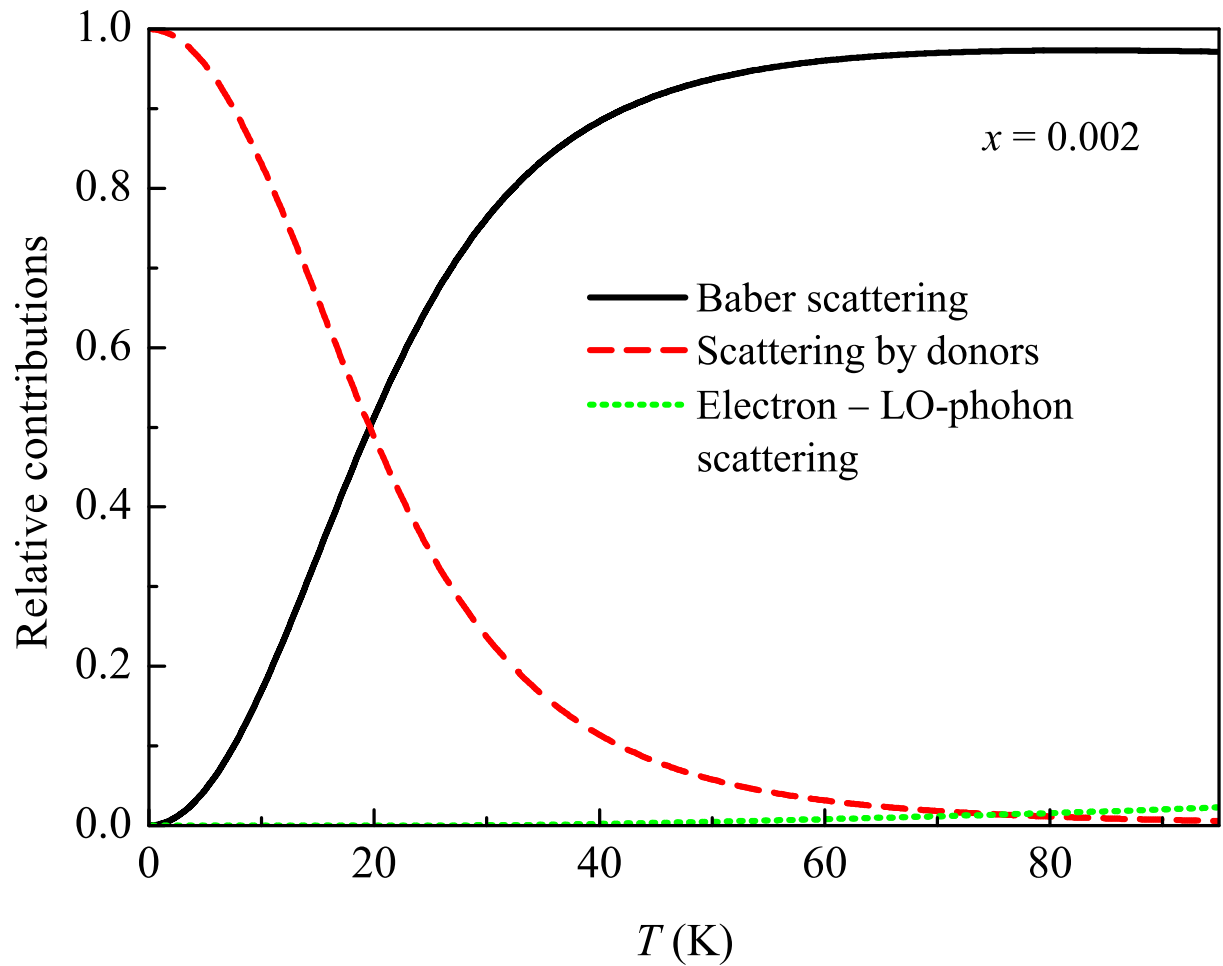}%
\caption{Relative contributions of different scattering mechanisms to the
total relaxation rate as a function of temperature for the doping content
$x=0.002$.}%
\end{center}
\end{figure}

\newpage%

\begin{figure}
[ph]
\begin{center}
\includegraphics[
height=5.0211in,
width=6.3373in
]%
{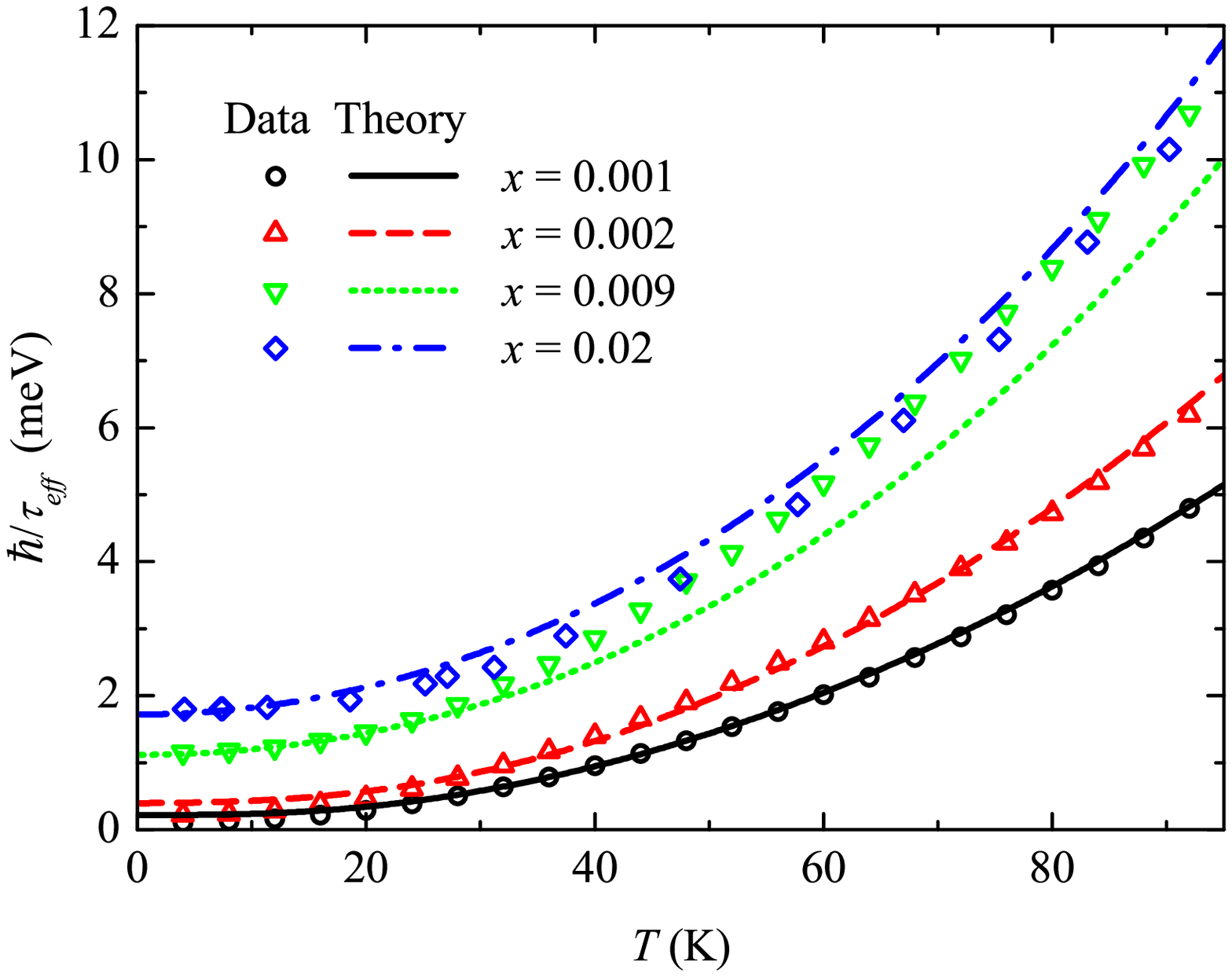}%
\caption{\emph{Curves}: the calculated total relaxation rate in Nb-doped
SrTiO$_{3}$ as a function of the temperature for the actual electron densities
corresponding to the experimental values of the doping \cite{VDM2011}.
\emph{Symbols}: experimentally determined relaxation rate in Nb-doped
SrTiO$_{3}$ from Fig. 2 of Ref. \cite{VDM2011}.}%
\end{center}
\end{figure}


\begin{thebibliography}{99}                                                                                               %


\bibitem[a]{altaff1}On leave of absence from: Department of Theoretical
Physics, State University of Moldova, str. A. Mateevici 60, MD-2009 Kishinev,
Republic of Moldova.

\bibitem[b]{altaff2}Also at Lyman Laboratory of Physics, Harvard University,
Cambridge, MA 02138, USA.

\bibitem {VDM2011}D. van der Marel, J. L. M. van Mechelen, and I. I. Mazin,
Phys. Rev. B \textbf{84}, 205111 (2011).

\bibitem {PRB-STO}J. T. Devreese, S. N. Klimin, J. L. M. van Mechelen, and D.
van der Marel, Phys. Rev. B \textbf{81}, 125119 (2010).

\bibitem {Koonce}C. S. Koonce, M. L. Cohen, J. F. Schooley, W. R. Hosler, and
E. R. Pfeiffer, Phys. Rev. \textbf{163}, 380 (1967).

\bibitem {Reyren}N. Reyren, S. Thiel, A. D. Caviglia, L. F. Kourkoutis, G.
Hammerl, C. Richter, C. W. Schneider, T. Kopp, A.-S. Ruetschi, D. Jaccard, M.
Gabay, D. A. Muller, J.-M. Triscone, and J. Mannhart, Science \textbf{317},
1196 (2007).

\bibitem {Lawrence}W. E. Lawrence and J. W. Wilkins, Phys. Rev. \textbf{7},
2317 (1973)

\bibitem {McD1}A. H. MacDonald, Phys. Rev. Lett. \textbf{44}, 489 (1980).

\bibitem {McD2}A. H. MacDonald, R. Taylor, and D. J. W. Geldart, Phys. Rev. B
\textbf{23}, 2718 (1981).

\bibitem {VDM-PRL}J. L. M. van Mechelen, D. van der Marel, C. Grimaldi, A. B.
Kuzmenko, N. P. Armitage, N. Reyren, H. Hagemann, and I. I. Mazin, Phys. Rev.
Lett. \textbf{100}, 226403 (2008).

\bibitem {TD2001}J. Tempere and J. T. Devreese, Phys. Rev. B \textbf{64},
104504 (2001).

\bibitem {Zhao}G. M. Zhao, V. Smolyaninova, W. Prellier, and H. Keller, Phys.
Rev. Lett. \textbf{84}, 6086 (2000).

\bibitem {Maebashi}H. Maebashi and H. Fukuyama, J. Phys. Soc. Jap.
\textbf{66}, 3577 (1997).

\bibitem {Baber}W. G. Baber, Proc. R. Soc. A \textbf{158}, 383 (1937).

\bibitem {Giamarchi}T. Giamarchi and B. S. Shastry, Phys. Rev. B \textbf{46},
5528 (1992).

\bibitem {Ziman}J. M. Ziman, \emph{Electrons and Phonons} (Oxford U.P.,
Oxford, England, 1960).

\bibitem {Mahan}G. D. Mahan, \emph{Many-Particle Physics}, second edition
(Plenum Press, 1990).

\bibitem {PD1985}F. M. Peeters and J. T. Devreese, Phys. Rev. B \textbf{32},
3515 (1985).

\bibitem {Bell}R. O. Bell and G. Rupprecht, Phys. Rev. \textbf{129}, 90 (1963).

\bibitem {Anselm}A. I. Anselm, \emph{Introduction to Semiconductor Theory}
(Mir, Moscow/Prentice-Hall, Englewood Cliffs, NJ, 1981).

\bibitem {Landau3}L. D. Landau and L. M. Lifshitz, \emph{Quantum Mechanics
Non-Relativistic Theory}, Vol. 3.

\bibitem {Luo}Y. Li, X. Luo and H. Kr\"{o}ger, Sci. China G: Phys. Mech. and
Astr. \textbf{49}, 60 (2006).
\end{thebibliography}
\end{document}